\documentclass[
    ,final            
  ]
  {aipproc}

\newcommand\apj{ApJ}                 
\newcommand\aap{A\&A}            

\newcommand\pra{Phys. Rev. A.}

\newcommand\physrep{Physics Reports}
\newcommand\nat{Nature}
\newcommand\Omm{\Omega_{\mbox{\rm \small m}}}
\newcommand\Omtot{\Omega_{\mbox{\rm \small tot}}}

\layoutstyle{8x11single}

\begin{document}

\title{A Hint of Poincar\'e Dodecahedral Topology in the WMAP First Year Sky Map}


\author{Boudewijn F. Roukema}{
  address={Toru\'n Centre for Astronomy, N. Copernicus University,
ul. Gagarina 11, PL-87-100 Toru\'n, Poland}
}

\begin{abstract}
Several analyses of the cosmic microwave background map made by the
satellite WMAP suggest that the global shape of a spatial section of
the Universe is that of a Poincar\'e dodecahedral space. A summary of
some of these analyses and a description of independent tests which
should be able to either increase confidence in the hypothesis or else
refute it to extremely high significance will be presented.
\end{abstract}

\maketitle


\section{Cosmic topology}

The possibility of the Universe (or more carefully, of a comoving spatial
section of the Universe) being {\em multiply connected} is first
known to have been suggested by \citet{Schw00,Schw98}, prior to the 
theories of special and general relativity which showed that there is 
no mathematically fundamental requirement for the {\em curvature} of 
the Universe to be zero.

General relativity provided a physical theory tightly relating
{\em curvature} to matter-energy density.
However, no clear theory relating 
{\em topology} to other physical constituents of the Universe has so far
been proposed (though some hints are starting to surface; see 
\citet{UzanTwins02} and \citet{LevinTwins01} for the relation between 
multiple connectedness and the comoving reference frame; and \citet{RBBSJ06} 
for a link between multiple connectedness and dark energy). For this reason,
most observational cosmology research into Universe geometry has concentrated
on empirically constraining the parameters of the perturbed 
Friedman-Lema\^{\i}tre-Robertson-Walker (FLRW) solution to the Einstein
field equations.

However, in the last decade, research has accelerated into investigating 
the (possibly not measurable) global shape of the Universe, i.e. its topology.
For recent reviews of this work, see
\citet{LaLu95}, \citet{Lum98}, \citet{Stark98}, \citet{LR99};
workshop proceedings: \citet{Stark98} and following articles,
and \citet{BR99}; and for detection strategies including both two-dimensional
methods (based on temperature fluctuations in the surface of last
scattering) and three-dimensional methods (based on distributions
of gravitationally collapsed objects distributed in three-dimensional
comoving space) see \citet{ULL99b,LR99,Rouk02topclass,RG04}.

\section{WMAP analyses: the Poincar\'e Dodecahedral Space}


The WMAP first-year observational all-sky map of the cosmic microwave background 
has generated considerable interest due to the weakness of large-scale
correlations, i.e. the amplitudes of the low-$l$ spherical harmonic 
components of the map are unexpectedly low and they seem to be
aligned (this has been referred
to as ``the axis of evil'').

Several different groups have now found that the multiply connected 
Poincar\'e dodecahedral space (PDS) better models the statistics of 
the observed WMAP map than simply connected, infinite, flat models:
\citet{LumNat03,Aurich2005a,Aurich2005b,Gundermann2005}.  Both classes of
models 
assume a standard FLRW solution to the Einstein field equations.

\citet{RLCMB04}, again for the FLRW metric, used the identified circles principle 
\citep{Corn96,Corn98b} to test the PDS hypothesis in the WMAP data and
found the most correlated circles appear for circle 
radii of $\alpha =11\pm1^{\circ}$ ,
for a left-handed screw motion when matching opposite circles, 
but not for a right-handed one, nor 
for an unphysical, control, zero rotation. 
The favoured six dodecahedral face centres and circle radii are listed
in Table~\ref{t-dodec}.
These six pairs of circles {\em independently} each
favour a circle angular radius of $11\pm1^{\circ}$. 
The temperature fluctuations along the matched circles are shown in Figs~13--17
of \citet{RLCMB04} and are clearly highly correlated.

{\begin{table}
\begin{tabular}{c c c c} \hline 
\rule{0ex}{2.5ex}
$i$ &
$l^{II}$ \mbox{ in $^{\circ}$} & $b^{II}$ \mbox{ in $^{\circ}$} 
& $\alpha$ \mbox{ in $^{\circ}$}\\ \hline 
   1 &     252.4 &  64.7    & 9.8 \\
   2 &      50.6  & 50.8   & 10.7 \\
   3 &     143.8  & 37.8 & 10.7 \\
   4 &     207.5  & 9.5 &   10.7 \\
   5 &     271.0  & 2.7 &  11.8 \\
   6 &     332.8   & 25.0 &    10.7 \\
\hline
\end{tabular}
\caption{Galactic sky coordinates of the six face centres for the 
(positively curved) dodecahedron
which shows excess values of the correlation statistic $S$ in 
\protect\citet{RLCMB04}, listing 
the face number $i$, galactic
longitude, latitude and estimated circle radius $\alpha$ (all in degrees).
The other 6 faces are directly opposite to the ones listed.
The orientation of the $36^{\circ}$ screw motion used to match 
faces is left-handed.
\label{t-dodec}}
\end{table}
}  

\section{Independent tests}

\subsection{CMB data: phase tests}

The \citeauthor{RLCMB04} solution can be tested by several consistency
tests, including phase tests. If the solution is the correct one, then 
it should mostly likely be an optimum in phase space in several different
directions. For example, allowing for an arbitrary (non-physical) phase
of rotation between matched circles should imply an optimal phase of 
$-36^{\circ}$.

\begin{figure}
  \includegraphics[height=.4\textheight]{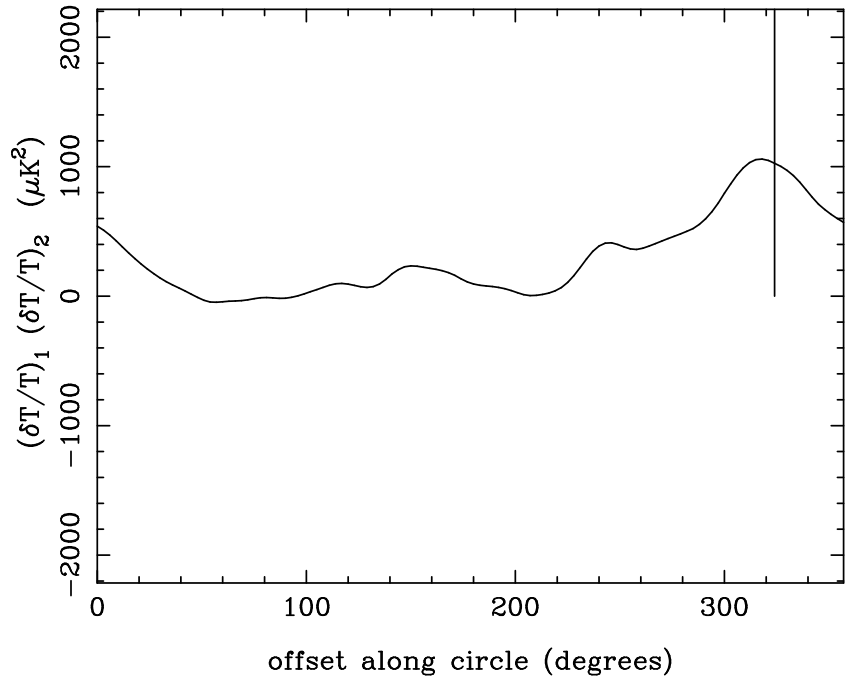}
  \caption{Mean temperature correlation in $\mu K^2$, 
for five of the six pairs of the \protect\citet{RLCMB04} PDS solution, 
as a function of offset phase between the circles, for the Integrated Linear
Combination (ILC) map of the WMAP first-year data. 
Circle pair \#4 is excluded here
because large parts of the circles fall on the galactic plane, so that a phase
test is of low significance. 
The expected phase of $-36^{\circ}$ is shown with a vertical line. 
Note that correlations are {\em not} constrained to be positive. It is 
reasonable that the random properties of the data lead to 
correlations at the non-optimal phases which mostly
cancel to what happen to be small positive values. Only at the expected
phase do these correlations add up to a {\em large} positive value.
\label{f-phase}}
\end{figure}

Fig.~\ref{f-phase} shows that the solution is clearly consistent with the
optimal phase.

\subsection{Total density parameter $\Omtot$}

The tests by 
\citet{LumNat03,Aurich2005a,Aurich2005b,Gundermann2005} require 
assumptions regarding the power spectrum of density perturbations and
the gaussianity or non-gaussianity of the distributions of amplitudes of the
perturbations.

Nevertheless, both these analyses and the \citet{RLCMB04} analysis make
strong, highly falsifiable predictions for $\Omtot$, which must be strictly
greater than unity for the PDS hypothesis to be correct.

\citet{RLCMB04} predict $\Omtot= 1.009\pm0.001$ (1$\sigma$) 
for $\Omm = 0.28\pm0.02$; 
the other groups predict slightly higher values.

\citet{Eisenstein05}, using a standard ruler method from the Sloan Digital 
Sky Survey, as in \citet{RMB02}, found $\Omtot \approx 1.015\pm 0.015$.

If future tests eventually found, for example, 
$\Omtot \approx 1.001\pm 0.001$, then the PDS hypothesis (whether from statistical
analyses or identified circles analysis) would be excluded to high significance.

\subsection{Other tests}

Other tests which avoid making assumptions on hypothetical statistical ensembles 
of universes include:

\begin{list}{$\bullet$}{\usecounter{enumi}}
\item separating na{\"{\i}}ve-SW, ISW and doppler components
\item foreground ``predictions''
\item polarisation data --- see \citet{Riaz06polar}
\end{list}

\section{GPL software}

The reader may carry out circles tests by using and/or modifying and/or
redistributing (under the terms of the GPL licence) the {\sc circles} 
package available at: 
\url{http://cosmo.torun.pl/GPLdownload/dodec/}.

Installation and example usages include:

\begin{list}{$\bullet$}{\usecounter{enumi}}
\item  {\em ./configure \&\& make \&\& make install}
\item  {\em circles -{}-help; info circles; man circles}
\item  {\em circles -{}-statistics} \\ (correlation calculations)
\item  {\em circles -{}-circles} \\ (plot the circles)
\item  {\em circles -{}-plot-phase} \\ (phase plots)
\item  short form: \\
{\em circles -s -c -P -d /scratch/topowork}  \\
do everything and use data files in /scratch/topowork/.
\end{list}

The data files are available as:
\begin{list}{$\bullet$}{\usecounter{enumi}}
\item \url{http://lambda.gsfc.nasa.gov/data/map/ilc/map_ilc_yr1_v1.fits}
\\ --- the WMAP ILC map
\item \url{http://cosmo.torun.pl/WMAPdata} 
\\  --- {\em secondary} files for default
installation (in principle, these should not be necessary, but as of
circles-0.1.27, it will be easier if the user downloads these.)
\end{list}



\bibliographystyle{aipproc}   





\end{document}